\documentclass[12pt,a4paper]{article}
\usepackage{latexsym,amsmath,amssymb,amsfonts,epic,oldgerm,eufrak}
\newcommand{\bea}{\begin{eqnarray}}
\newcommand{\eea}{\end{eqnarray}}
\newcommand{\beq}{\begin{equation}}
\newcommand{\eeq}{\end{equation}}
\newcommand{\iden}{\mbox{{1}\hspace{-.11cm}{l}}}
\newcommand{\nn}{\nonumber}
\newcommand{\noi}{\noindent}
\newcommand{\bs}{\begin{split}}
\newcommand{\es}{\end{split}}
\def\a{\alpha}
\def\ap{\alpha'}
\def\b{\beta}

\def\d{\delta}
\def\D{\Delta}
\def\di{\displaystyle}
\def\e{\epsilon}

\def\g{\gamma}
\def\G{\Gamma}

\def\im{-\infty}
\def\ip{+\infty}
\def\k{\kappa}
\def\l{\lambda}

\def\s{\sigma}
\def\sx{\left}
\def\m{\mu}
\def\n{\nu}

\def\o{\omega}
\def\O{\Omega}

\def\ot{\otimes}
\def\p{\pi}

\def\der{\partial}

\def\dx{\right}
\def\s{\sigma}

\def\ve{\varepsilon}

\begin{document}
\begin{titlepage}

\vskip 0.5cm

\hfill\hbox{\today}

\vskip 2cm

\begin{center}

{\LARGE\bf {Penrose limit of $AdS_3 \otimes S^3 \otimes S^3 \otimes S^1$ and 
its\\ associated $\s$-model}}\\

\vskip 1.5cm

{\bf Luca Sommovigo} \\

\vskip 0.5cm 

{\small \sl Dipartimento di Fisica, Politecnico di Torino,\\
 Corso Duca degli Abruzzi 24, I-10129 Torino\\
and Istituto Nazionale di Fisica Nucleare (INFN) - Sezione di
Torino, Italy\\ luca.sommovigo@polito.it}\\

\vspace{6pt}

\end{center}

\vskip 3cm

\begin{abstract}

We study the Penrose limit of a supersymmetric IIB background with non-trivial
NS 3-form field strength, obtaining a solution with the smallest number of 
supercharges (i.e. 16) allowed; we write down explicitly the superalgebra 
of the theory, build the supersymmetric associated IIB string $\s$-model 
 and make conjectures on the dual gauge theory.

\end{abstract}

\vfill

\end{titlepage}
\section{Introduction}
Since the middle 70's, it is believed that large N gauge theories have a 
string theory description, even if these strings live in a higher dimensional 
space than gauge theory; this idea stands as the basis of the so called 
$AdS/CFT$ correspondence (for an exhaustive review, see 
\cite{Aharony:1999ti}), where a conformal field theory ($CFT$) in $(p+1)$ 
dimensions is related to superstring theory in $AdS_{p+2} \otimes X$, with 
$X$ a generic Einstein compact manifold.\\
Unfortunately, it is not possible to study the complete spectrum of the 
string theory in any arbitrary background, due to technical difficulties in 
quantization, and one is obliged to study it only for large values of 
the radius of both $AdS$ and $X$ , that is in a supergravity approximation; 
but, being $R \sim \sx(g_{YM}^2 N\dx)^\frac{1}{4}$, this means that only the 
sector of the $SCFT$ with large 't Hooft coupling is involved.\\
Going beyond this limit is essential, for instance, in order to find
 quantitative 
results from string theory for the large N limit of gauge theories with 
finite 't Hooft coupling and also to prove (and not only 
to conjecture) the $AdS/CFT$ correspondence.\\
But actually $AdS$ is not the simplest supergravity background: Blau, 
Figueroa-O'Farrill, Hull and Papadopoulos \cite{Blau:2001ne} 
demonstrated that there exists another maximally supersymmetric background in 
IIB supergravity, which can be obtained as the Penrose limit of the 
$AdS_5 \otimes S_5$ background, and whose metric is non-flat only in two 
directions:

\begin{equation}
\di ds^2 = 2 du dv + \sum_{A,B=1}^{8}\left(\mathcal{A}^{AB}~y_A y_B\right) 
du^2 + \sum_{A=1}^{8} \left( dy_A \right)^2 \nn
\end{equation}

\noindent with a non-trivial 5-form:

\begin{equation}
F_{u1234} = F_{u5678} = \mathrm{const.} \nn
\end{equation}

Subsequently, it was shown \cite{Metsaev:2001bj,Metsaev:2002re} that these 
types of background lead to a superstring $\sigma$-model that is easily 
quantized in the light-cone gauge, since it is quadratic in all the fields.  
The next important step in this direction was achieved by Berenstein, 
Maldacena and Nastase (BMN) \cite{Berenstein:2002jq}, who extended the 
$AdS/CFT$ duality to pp-wave background in string theory, by claiming that 
the string theory in a background obtained from $AdS_5 \otimes S_5$ after a 
Penrose limit should be dual to the large N limit of a certain sub-sector of 
four dimensional $\mathcal{N}$=4 SYM.\\
This sub-sector is characterized by 
choosing a $U(1)_R$ subgroup of the usual $SU(4)_R$ R-symmetry group of the 
gauge theory and by considering states with conformal weight $\Delta$ and 
R-charge J scaling as $\sqrt{N}$, so that both of them are large, but 
with $\Delta - J$ finite.\\
After \cite{Berenstein:2002jq}, the Penrose limit was applied to other 
models, with different amount of supersymmetries and different geometries, 
and some 
results were obtained about the non-perturbative behaviour of a large variety 
of gauge theories; in particular, 
in \cite{Hikida:2002in} superstrings on the Penrose limits of $AdS_3 \ot S^3 
\ot T^4$ and $AdS_3 \ot S^3 \ot K^3$ have been investigated; 
in \cite{Gomis:2002qi}, BMN duality is 
tested when the superstring background is $AdS_3 \otimes S_3 \otimes K^3$, 
finding an agreement between CFT particle and string spectra at 
least in the leading order in $\frac{J^2}{N}$, 
while in \cite{Lunin:2002fw} a map between $CFT$ operators and string 
oscillators is found in an $AdS_3 \ot S^3 \ot T^4$ background; 
in \cite{Cho:2002zp}, the Penrose limit of 
$AdS_3 \otimes S^3 \otimes T^4$ is studied together with its supersymmetry 
content, and in \cite{Russo:2002rq} solutions of the type $AdS_3 \otimes S_3$ 
are studied with either NS-NS or R-R 3-form background. \\
Motivated by these 
considerations, and moving from our precedent work where we studied 
compactifications of IIB supergravity on generic backgrounds of the type 
$AdS_3 \otimes G/H$ \cite{Castellani:2000nb}, in this paper we study the 
Penrose limit of a IIB configuration with non-trivial NS 3-form, its 
supersymmetry content and the supersymmetric $\s$-model Hamiltonian in this 
background; the paper is organized as follows: in section 
\ref{pl} and \ref{geo} we perform the Penrose limit of our supergravity 
background and calculate the spin connection, the Riemann and the Ricci 
tensors. In section \ref{ss}, we study the
supersymmetry variations of gravitino and dilatino, and  
show that there is no supersymmetry enhancement; moreover, the 
number of supersymmetry charges conserved is the minimum for a pp-wave 
background, and this leads to the fact that bosons and fermions in the $\s$-
model have different masses (that is, the supersymmetry is completely broken 
in the light-cone gauge \cite{Cvetic:2002nh}).\\
In section \ref{superalgebra} we calculate explicitly the supersymmetry 
algebra: the bosonic sector is obtained as an In\"on\"u-Wigner contraction 
of the parent algebra, while the mixed and fermionic sectors are obtained
by means of the spinorial Lie derivative and of the structure of the Killing 
spinors.\\ 
In section \ref{lagr} we write down the supersymmetric $\s$-model Lagrangian 
both in bosonic and in fermionic sectors, and we can observe another 
interesting feature of this model: as a remnant of the 
parent background, where the radii of $AdS$ and $S$ were different, 
the eight bosons are divided into three sets: four of them are massless, 
coming from the $S^3 \otimes S^1$ part trivially involved in the limit, 
two have mass $\frac{1}{2} \mu^2$ and the others $\frac{1}{4} \mu^2$, being 
respectively the embedding fields coming from coordinates of $AdS_3$ and 
$S^3$. \\
In section \ref{hamilton}, we compute explicitly the supersymmetric 
$\sigma$-model Hamiltonian and evaluate its spectrum, which is in perfect 
agreement with that found by BMN, where:

\bea
& H_{string} = \o_n \mathcal{N}_n \nn\\
& \omega_n = \sqrt{\mu^2 + \frac{n^2}{(\alpha' p^+)^2}}\nn
\eea

\noindent with the only exception for an interaction term that generates an 
extra piece in the eigenvalues of the string Hamiltonian:

\begin{equation}
\omega_n = \sqrt{a \sx(\mu p^+ \ap\dx)^2 \pm b \sx(\mu p^+ \ap\dx)n + n^2}\nn
\end{equation}

\noindent where $a$ and $b$ are constants.\\
Finally, section \ref{maldacena} contains some comments about the 
$SCFT$ dual to the parent background and about BMN duality in this case.

\section{Penrose limit of $AdS_3 \otimes S_3 \otimes S_3 \otimes S_1$} 
\label{pl}

Let us start with a particular solution of type IIB supergravity, 
$AdS_3 \otimes S_3 \otimes S_3 \otimes S_1 $ \cite{Castellani:2000nb}, 
with nonzero 3-form $G$, whose metric reads:

\beq
\label{dsquadro}
\begin{array}{c}
ds^2 = R^2 [ \cosh^2 \rho ~dt^2 - d\rho^2 - \sinh^2 \rho ~d\alpha^2 - 2 ( 
\cos^2 \theta_1~ d\psi_1^2 + d\theta_1^2 + \sin^2 \theta_1~ d\phi_1^2 + \\
+ \cos^2 \theta_2~ d\psi_2^2 + d\theta_2^2 + \sin^2 \theta_2~ d\phi_2^2 + 
d \theta^2 )] 
\end{array}
\eeq

\noindent where $\rho$, $\alpha$ and $\tau$ are global coordinates (see 
appendix \ref{adsandsphere}) on $AdS_3$, whose radius is $R$; $\theta_{1,2}$, 
$\psi_{1,2}$ 
and $\phi_{1,2}$ are coordinates of the two 3-spheres, whose radius is 
$\sqrt{2}R$, and $\theta$ parametrizes $S_1$; finally, $G$ is defined by:

\beq
\label{Gform}
\begin{array}{c}
G = 2\sqrt{2} R^2 \cosh \rho ~ \sinh \rho ~dt~ d\rho ~d\alpha + 4 \sqrt{2} 
R^2 \sin \theta_1 ~\cos \theta_1 ~d\psi_1 ~d\theta_1 ~d \phi_1+ \\
+ 4 \sqrt{2} R^2 \sin \theta_2 ~\cos \theta_2~ d\psi_2 ~d\theta_2 ~d \phi_2. 
\end{array}
\eeq

\noindent Notice that, in the light of BMN correspondence, it is convenient 
to use global coordinates because in this way the space-time energy of a 
string state is directly related to the conformal dimension of the 
corresponding CFT state in dual gauge theory. In order to study the pp-wave 
configuration, we need to change coordinates from (\ref{dsquadro}) to:

\beq
\label{change}
\begin{array}{c}
\di t = \frac{\mu}{\sqrt{2}} u + \frac{\omega^2}{\sqrt{2} \mu R^2} v \\
\di \psi_1 = \frac{\mu}{2} u - \frac{\omega^2}{2 \mu R^2} v \\
\di \rho = \frac{\omega}{R} y_{12} \\
\di \theta_1 = \frac{\omega}{\sqrt{2}R} y_{34} \\
\di \alpha = \arctan {\frac{y_2}{y_1}} \\
\di \phi_1 = \arctan {\frac{y_4}{y_3}} \\
\di \theta_2 = \frac{\omega}{\sqrt{2}R} y_{56} \\
\di \phi_2 = \arctan {\frac{y_6}{y_5}} \\
\di \psi_2 = \frac{\omega}{\sqrt{2}R} y_7 \\
\di \theta = \frac{\omega}{\sqrt{2}R} y_8 
\end{array}
\eeq

\noindent where $y_{AB} = \left( {y_A}^2 + {y_B}^2\right)^{\frac{1}{2}}$ 
($A$,$B$ run over $1, \dots ,8$), $\omega$ is a rescaling dimensionless 
parameter and $\mu$ a dimensional parameter which will be interpreted as 
mass in the light-cone gauge $\sigma$-model Lagrangian; these coordinates 
are well suited to describe the geometry before taking the limit, since they 
describe the space from the viewpoint of the null geodesic ($\omega$ 
represents the distance from the null geodesic in units of $R$). Using these 
definitions in (\ref{dsquadro}), we rewrite the metric as function of the 
new variables:

\beq
\begin{array}{c}
\di \frac{ds^2}{R^2} = \frac{\mu^2}{2}\left(\cosh^2 \frac{\omega}{R} y_{12} - 
\cos^2 \frac{\omega}{\sqrt{2} R}y_{34}\right)du^2 + \frac{\omega^4}{2 \mu^2 
R^4} \left(\cosh^2 \frac{\omega}{R} y_{12} + \cos^2 \frac{\omega}{\sqrt{2} R}
y_{34} \right)dv^2 \\
\di + \frac{\omega^2}{R^2}\left(\cosh^2 \frac{\omega}{R} y_{12} + \cos^2 
\frac{\omega}{\sqrt{2} R}y_{34} \right)dudv - \left(\frac{\omega^2}{R^2}
\frac{{y_1}^2}{y_{12}^2} + \frac{{y_2}^2}{y_{12}^4} \sinh^2 \frac{\omega}{R} 
y_{12} \right){dy_1}^2 + \\
\di - \left(\frac{\omega^2}{R^2}\frac{{y_2}^2}{y_{12}^2} + 
\frac{{y_1}^2}{y_{12}^4} \sinh^2 \frac{\omega}{R} y_{12} \right){dy_2}^2 - 2 
\frac{y_1y_2}{y_{12}^4}\left(\frac{\omega^2}{R^2}y_{12}^2 - \sinh^2 
\frac{\omega}{R} y_{12}\right)dy_1dy_2 +\\
\di - \left(\frac{\omega^2}{R^2}\frac{{y_3}^2}{y_{34}^2} + 2 
\frac{{y_4}^2}{y_{34}^4} \sin^2 \frac{\omega}{\sqrt{2}R} y_{34} \right)
{dy_3}^2 -  \left(\frac{\omega^2}{R^2}\frac{{y_4}^2}{y_{34}^2} + 2 
\frac{{y_3}^2}{y_{34}^4} \sin^2 \frac{\omega}{\sqrt{2}R} y_{34} \right)
{dy_4}^2 + \\
\di - 2 \frac{y_3y_4}{y_{34}^4}\left(\frac{\omega^2}{R^2}y_{34}^2 - 2 \sin^2 
\frac{\omega}{\sqrt{2}R} y_{34}\right)dy_3dy_4 - \left(\frac{\omega^2}{R^2}
\frac{{y_5}^2}{y_{56}^2} + 2 \frac{{y_6}^2}{y_{56}^4} \sin^2 
\frac{\omega}{\sqrt{2}R} y_{56} \right){dy_5}^2 + \\
\di - \left(\frac{\omega^2}{R^2}\frac{{y_6}^2}{y_{56}^2} + 2 
\frac{{y_5}^2}{y_{56}^4} \sin^2 \frac{\omega}{\sqrt{2}R} y_{56} 
\right){dy_6}^2 - 2 \frac{y_5y_6}{y_{56}^4}\left(\frac{\omega^2}{R^2}
y_{56}^2 - 2 \sin^2 \frac{\omega}{\sqrt{2}R} y_{56}\right)dy_5dy_6 + \\
\di - \frac{\omega^2}{R^2} \cos^2 \frac{\omega}{\sqrt{2}R} y_{56}~ {dy_7}^2 -
\frac{\omega^2}{R^2}~{dy_8}^2 
\end{array}
\eeq

Using now the definition of Penrose limit for the metric, i.e. $\di 
ds^2_\o = \lim_{\o \to 0} \frac{1}{\o^2} ds^2(\o)$, we get:

\begin{equation}
ds^2_{\omega} = \frac{1}{2} \mu^2 \left( y_{12} + \frac{1}{2} y_{34} 
\right) du^2 + 2 du dv - dy_A dy_A. \label{newmetric}  
\end{equation}

We could have expected a similar result because geometrically speaking the 
Penrose limit enlarges and stretches a portion of the parent space near 
a null geodesic (in this case identified by 
$\rho = \psi_2 = \theta_1 = \theta_2 = \theta = 0$ with affine parameter 
$u$), the mathematical effect being to unwrap the angular coordinates 
orthogonally to the $u-v$ plane, making the sphere geometry flat. \\
The same procedure must be performed \cite{Gueven:2000ru} over the 3-form 
(\ref{Gform}), that is the only nonzero field; after the change of 
coordinates, we get:

\beq
\begin{array}{c}
\di G_{u12} = 2 \mu \omega R ~\frac{1}{y_{12}}~ \cosh \frac{\omega}{R}y_{12} ~
\sinh\frac{\omega}{R}y_{12} \\
\di G_{v12} = \frac{2\omega^3}{\mu R}~\frac{1}{y_{12}}~\cosh\frac{\omega}{R}
y_{12} ~\sinh\frac{\omega}{R}y_{12} \\
\di G_{u34} = 2 \mu \omega R~\frac{1}{y_{34}} ~\sin\frac{\omega}{\sqrt{2}R} 
y_{34} ~\cos \frac{\omega}{\sqrt{2}R} y_{34} \\
\di G_{v34} = \frac{2\omega^3}{\mu R}~\frac{1}{y_{34}} ~\sin
\frac{\omega}{\sqrt{2}R} y_{34} ~\cos \frac{\omega}{\sqrt{2}R} y_{34} \\
\di G_{567} = - 2 \sqrt{2} \omega^2 ~\frac{1}{y_{56}}~\sin
\frac{\omega}{\sqrt{2}R} y_{56} ~\cos\frac{\omega}{\sqrt{2}R} y_{56} 
\end{array}
\eeq

\noindent and, since $\di G^\o = \lim_{\o \to 0} \frac{1}{\o^2} G(\o)$, 

\beq
\label{newG}
\begin{array}{c}
G^{\omega}_{u12} = 2 \mu \\
G^{\omega}_{u34} = \sqrt{2} \mu 
\end{array}
\eeq

\noindent while all other components vanish.
It is important to notice that this background satisfies the equations of 
motion of IIB supergravity because of the so called hereditary property of 
the Penrose limit \cite{Blau:2002mw}. Another remark is in order: the radius 
of $S^1$ is not really fixed by the equations of motion: so that in 
(\ref{newmetric}) we have chosen it to coincide with the last $S^3$ radius 
even if it is not necessary; this fact will become evident later, when we 
will study the superalgebra of the background.

\section{Geometry of pp-wave solution}
\label{geo}

In studying the geometry of configuration (\ref{dsquadro}), it 
is convenient to use flat space light cone indices (that is 
$E^{\pm} = \frac{1}{2} \left( E^0 \pm E^9\right)$), the flat metric being 
(again, $A,B:1,\dots,8$):

\begin{equation}
\eta_{++} = \eta_{--} = 0~;~\eta_{+-} = 1~;~\eta_{AB} = -\delta_{AB} 
\end{equation}

\noindent the vielbein ($E^M_\mu$) of our background is:

\beq
\begin{array}{c}
  E^+_u = 1 \\
  E^+_v = 0 \\
  E^-_u = \frac{1}{4} \mu^2 \left(y_{12}^2 + \frac{1}{2} y_{34}^2 \right)\\
  E^-_v = 1 \\
  E_A^A = 1 
\end{array}
\eeq

\noindent and its inverse:

\beq
\begin{array}{c}
  E^u_+ = 1\\
  E^u_- = 0\\
  E^v_+ = -\frac{1}{4} \mu^2 \left( y_{12}^2 + \frac{1}{2} y_{34}^2 \right)\\
  E^v_- = 1\\
  E^A_A = 1. 
\end{array}
\eeq

As a consequence, in flat indices, $G$ takes the values:

\beq
\begin{array}{c}
  G^{\omega}_{+12} = 2 \mu \\
  G^{\omega}_{+34} = \sqrt{2} \mu.
\end{array}
\eeq

By means of Cartan-Maurer equations implemented with null-torsion condition 
($d E^M + \omega^M_{~N} \wedge E^N = 0$) , we can write down the Riemannian 
connection:

\beq
\label{spinconn}
\begin{array}{c}
  \omega_{u~+i} = \frac{1}{2} \mu^2 y_i \\
  \omega_{u~+m} = \frac{1}{4} \mu^2 y_m 
\end{array}
\eeq

\noindent and subsequently the Riemann tensor, defined as 
$\mathcal{R}_{MN} = d \omega_{MN} + \omega_M^{~~P} \wedge \omega_{PN}$; 
finally, we obtain Ricci tensor (
$R_{MN} = \mathcal{R}^P_{~MPN}$), whose only non zero component is:

\begin{equation}
R_{++} = \frac{3}{2} \mu^2, \nn 
\end{equation}

from which we can conclude that the curvature scalar ($R=R_{MN}\eta^{MN}$) is 
zero, being $\eta^{++}=0$.

\section{Supersymmetry}
\label{ss}

It is well known that any pp-wave configuration preserves, at least, all the 
supersymmetries of its parent background, as shown in 
\cite{Blau:2002mw}; this means that, since our parent background 
configuration ($AdS_3 \otimes S_3 \otimes S_3 \otimes S_1$ with nonzero 
3-form $G$) preserves one half of supersymmetry charges, the pp-wave 
configuration will preserve at least this same amount.
Requiring the vanishing of the dilatino supersymmetry variation brings us to 
the equation (from now on, we drop the label $\omega$ on the fields):

\begin{equation}
G_{MNP} \Gamma^{MNP} \epsilon =0 \label{dilatino}
\end{equation}

\noindent after some Dirac algebra, (\ref{dilatino}) can be 
translated into the condition:

\begin{equation}
\Gamma^+ \epsilon = \frac{1}{2} \left( \Gamma^0 + \Gamma^9 \right) 
\epsilon =0 \label{cond}
\end{equation}

\noindent which precisely halves the amount of supersymmetry charges preserved:
we can guess now that imposing the vanishing of the gravitino variation will
 not reduce any further the total amount of supersymmetry, but only determine 
the exact dependence of $\e$ on the coordinates; the explicit expression for 
$\d \psi_M=0$ is:

\begin{equation}
\nabla_M \epsilon + \frac{1}{96} \left(\Gamma^{~NPQ}_M G_{NPQ} - 9\Gamma^{PQ} 
G_{MPQ} \right) \epsilon^* =0\label{graviequ}
\end{equation}

\noindent where $\nabla_M$ is the standard covariant derivative (
$\nabla_M = \partial_M + \frac{1}{4} \omega_{M~NP} \Gamma^{NP}$).
In ten dimensions, spinors satisfy both Majorana and Weyl conditions: Weyl 
condition says that $\e$ has a definite chirality, and then only one half of 
its components are nonzero, while as a consequence of Majorana condition, we 
can easily separate the real from the imaginary part of equation 
(\ref{graviequ}); besides, after some algebra, we can rewrite the 
gravitino variation as follows:

\begin{equation}
\nabla_M \e^I + \s_3^{IJ}\frac{1}{96}\sx( \G_M \G^{NPQ} G_{NPQ} - 12 
\G^{PQ} G_{MPQ} \dx) \e^J =0
\end{equation}

\noindent where $I,J=1,2$ label the real and imaginary part of $\e$ (that is
 $\e = \e^1 + i \e^2$), and $\s_3$ is the usual third
Pauli matrix; the first supersymmetry condition (\ref{cond}) simplifies
equation (\ref{graviequ}), and we get: 

\begin{equation}
\left( \delta^{IJ} \nabla_M - \s_3^{IJ} \frac{1}{8} G_{MNP} 
~\G^{NP}\right) \epsilon^J = 0. \label{gravitino}
\end{equation}

Due to our background, in the directions from $5$ up to $8$, this condition 
simply reads:

\begin{equation}
\partial_q \epsilon^I = 0 \label{indonq}.
\end{equation}

Now we focus on $i$ and $m$ directions, where the Riemannian connection is 
always zero, while the $G$-term, after some manipulations, takes the form:

\beq
\begin{array}{c}
  G_{iMN}~\Gamma^{MN} = 2 \mu \Gamma_i W \Gamma^+ \\
  G_{mMN}~\Gamma^{MN} = \sqrt{2} \mu \Gamma_m Y \Gamma^+ 
\end{array}
\eeq

\noindent and the matrices $W$ and $Y$ are defined as:

\begin{equation}
W = \frac{1}{2} \varepsilon_{ij}~\Gamma^{ij} = \Gamma^1 \Gamma^2 ~,~ Y = 
\frac{1}{2} \varepsilon_{mn}~\Gamma^{mn} = \Gamma^3 \Gamma^4, \nn
\end{equation}

\noi recalling condition (\ref{cond}), we can rewrite equation 
(\ref{gravitino}) as:

\begin{equation}
\partial_i \e^I = \partial_m \e^I = 0. \label{imdirections}
\end{equation}

In the $-$ and $+$ directions, instead, we get:

\beq
\label{susy}
\begin{array}{c}
  \partial_- \epsilon^I = 0 \\
  \partial_+ \epsilon^I = \partial_u \epsilon^I = - \s_3^{IJ} \frac{1}{4}
 \mu \left( 2 W + \sqrt{2} Y \right) \epsilon^J 
\end{array} 
\eeq

\noindent differential equations (\ref{indonq}),(\ref{imdirections}) and 
(\ref{susy}) tell us that $\e$ is independent on $y_q$, $y_i$, $y_m$and $v$,
 while retaining the typical non trivial dependence on $u$, in fact the second 
of (\ref{susy}) can be easily integrated, being a linear first order 
differential equation with constant coefficients, and the explicit solution 
turns out to be:

\beq
\label{residualsusy}
\begin{array}{c}
\e(u)^I = \sx[ - \s_3^{IJ} \sin \sx(\frac{1}{2} \mu u \dx) W + \cos \sx(
\frac{1}{2} \mu u \dx) \iden \d^{IJ} \right] \times \\ 
\times \sx[ - \s_3^{IJ}\sin \sx( \frac{\sqrt{2}}{4} \mu u \dx) Y + \cos 
\sx( \frac{\sqrt{2}}{4} \mu u \dx) \iden \d^{IJ}\dx]\e(0)^J 
\end{array}
\eeq

\noindent where $\e(0)^I$ satisfies $\G^+ \e(0)^I= 0$; there is no 
supersymmetry enhancement: only one half of supersymmetry charges are 
conserved.\\
It is important to notice that, with a different ansatz, we could preserve 
a larger amount of supersymmetry: for instance in \cite{Lu:2002kw} 
a rotation of the two foliating circles within the two 3-spheres 
is taken, and supernumerary Killing spinors arise, when the rotation 
angle is $\a = \frac{\p}{4}$.  

\section{pp-wave solution superalgebra}
\label{superalgebra}

Any superalgebra $\EuFrak{G}$ can be divided into two subsets: a bosonic one, 
which we will refer to as $\EuFrak{G}_0$, and a fermionic one, $\EuFrak{G}_1$, 
spanned respectively by Killing vectors and spinors; both $\EuFrak{G}_0$ and 
$\EuFrak{G}_1$ will be evaluated starting from the generators of the isometry 
algebra of the parent background, following the method used in 
\cite{Blau:2002mw}.\\
In order to evaluate the even subalgebra, it is convenient to make a 
distinction, in the parent background, between the $AdS_3 \otimes S_3$ 
part, that is non-trivially modified by the Penrose limit, and 
$S_3 \otimes S_1$, where there is no coordinate mixing. The former subspace 
is invariant under the action of the algebra 
$\EuFrak{so}(2,2) \oplus \EuFrak{so}(4)$, which can be thought of as the 
algebra of rotations in the 8-dimensional space 
$\mathbb{R}^{2,2} \otimes \mathbb{R}^4$ where $AdS_3$ and $S_3$ are 
hyper-surfaces (see appendix \ref{adsandsphere}) \cite{Das:2002ij}; analogously 
the $S_3 \otimes S_1$ 
subspace is invariant under the algebra $\EuFrak{so}(4) \oplus \EuFrak{so}(2)$ 
of rotations in the 6-dimensional space $\mathbb{R}^4 \otimes \mathbb{R}^2$ 
where again $S_3$ and $S_1$ are hyper-surfaces.\\
In summary, the parent background can be viewed as a 10-dimensional 
hyper-surface lying in the space $\mathcal{M} = \mathbb{R}^{2,2} \otimes 
\mathbb{R}^4 \otimes \mathbb{R}^4 \otimes \mathbb{R}^2$, whose generators 
of isometries are:

\begin{equation}
\xi_{\Lambda \Sigma} = z_\Lambda \der_\Sigma - z_\Sigma \der_\Lambda \nn
\end{equation}      

\noi where $\der_\Lambda = \frac{d}{dz^\Lambda}$ and $z_\Lambda$, are 
cartesian coordinates of any of the four subspaces building blocks of the 
14-dimensional $\mathcal{M}$ space.\\
Actually, to describe the parent space geometry, we adopt global 
coordinates which are written in terms of the $z_\Lambda$'s in the appendix 
\ref{adsandsphere}.
Now (\ref{change}) is to be performed, and the generators turn out to be a 
complicated mixture of derivative operators multiplied by (hyperbolic) 
trigonometric functions of $u$ and $v$; before taking the limit, we must 
rescale the $\xi$'s by an appropriate power of $\omega$, which will be 
chosen so that all the limits exist and are non-zero: 

\begin{equation}
\di \xi^\o = \lim_{\o \to 0} \o^{\D_\xi} \xi,~ \D_\xi \in \mathbb{N} \nn
\end{equation}

\noindent this is a consequence of the fact, demonstrated by Geroch in 
\cite{Geroch:1969}, that, in a family of spaces labeled by a continuous 
parameter, if the number of Killing vectors is independent on the label, 
in any limit for particular values of the label the number of Killing vectors 
must remain unchanged, but these can describe a different algebra. 
This is precisely our case: $\omega$ parametrizes a continuous set of spaces, 
all solutions to IIB equations, characterized by the same number of Killing 
vectors (namely $6 + 6+ 6+ 1 =19$), then in the limit $\omega \rightarrow 0$ 
we must recover the same number of Killing vectors, but the algebra will no 
more be $\EuFrak{so}(2,2) \oplus \EuFrak{so}(4) \oplus \EuFrak{so}(4) \oplus 
\EuFrak{so}(2)$.\\
From a mathematical viewpoint those different rescalings correspond to an 
In\"on\"u-Wigner contraction on the parent algebra and lead to the 
following set of generators:

\beq
\begin{array}{c}
\di \xi_\pm = - \der_\pm \\
\di \xi_i = - y_i \frac{\mu}{\sqrt{2}} \cos \frac{\mu u}{\sqrt{2}} \der_- +\sin
\frac{\mu u}{\sqrt{2}} \der_i \\
\di \xi_m = - y_m \frac{\mu}{2} \cos \frac{\mu u}{2} \der_- + \sin 
\frac{\mu u}{2} \der_m \\
\di \xi_i^* = \cos \frac{\mu u}{\sqrt{2}} \der_i + y_i \frac{\mu}{\sqrt{2}} 
\sin \frac{\mu u}{\sqrt{2}} \der_-  \\
\di \xi_m^* = \cos \frac{\mu u}{2} \der_m + y_m \frac{\mu}{2}\sin 
\frac{\mu u}{2} \der_- \\
\di \xi_{ij}= - y_i \der_j + y_j \der_i \\
\di \xi_{mn}= - y_m \der_n + y_n \der_m \\
\di \xi_q = \der_q \\
\xi_{qr} = - y_q \der_r + y_r \der_q ~~(q,r~\neq~8). 
\end{array}
\eeq

An important comment is in order: as we said in section \ref{pl}, the $SO(4)$ 
isometry of the metric (\ref{newmetric}) is apparent: the invariance under 
rotations holds only in the directions labeled by ($5 \dots 7$).
Dealing now with the odd subalgebra, we can say that, since our solution 
preserves only one half of supersymmetry, $\EuFrak{G}_1$ has two 
generators, $Q^I,~I=1,2$, possessing 
only one half of the degrees of freedom of usual $SO(10)$ Majorana-Weyl 
spinors (they indeed satisfy the bound $\G^+ Q^I = 0$); the structure constants
 of their commutators with the even part of the supersymmetry algebra can be
 obtained using the spinorial Lie derivative, $\EuFrak{L}_\xi$ in the 
direction of the Killing vector $\xi$, 
which can be defined using the standard covariant derivative ($\nabla_M = 
\partial_M + \frac{1}{4} \omega_M^{~PN} \Gamma_{PN}$) as:

\begin{equation}
\EuFrak{L}_\xi \e^I = \xi^M \nabla_M^{IJ} \epsilon + \frac{1}{4} 
\nabla_{[M}^{IJ} \xi_{N]} \G^{MN} \e^J. \nn
\end{equation}

It can be shown that, by construction, $\EuFrak{L}_\xi$ preserves the space 
of Killing spinors i.e. let $\xi$ be a Killing vector field; acting on a 
generic Killing spinor $\e^I (\e^I_+,\e^I_-)$\footnote{Any spinor can 
be split in a $+$ and a $-$ part as follows: $\e = \eta^{+-}\e = \frac{1}{2}
\sx\{ \G^+,\G^- \dx\} \e = \frac{1}{2} \G^+ \G^- \e + \frac{1}{2} \G^- \G^+ 
\e$, so that $\G^\pm \e_\pm =0$ since ${\G^+}^2 = {\G^-}^2 = 0$}, 
$\EuFrak{L}_\xi$ will give  
a Killing spinor with different parameters $\e^I (S_\xi^+\e^I_+,S_\xi^- 
\e^I_-)$. This defines an action of the Lie algebra of isometries on the space
 of Killing spinors, whose structure constants are given by the  
matrices $S_\xi^+$ and $S_\xi^-$

\begin{equation}
\mathcal{L}_\xi \e^I = S_\xi^{IJ} \e^J. \nn
\end{equation}

Explicitly we can write:

\beq
\begin{array}{c}
\EuFrak{L}_+ \e^I = \frac{1}{4} \mu \sx( 2 W + \sqrt{2} Y \dx) \s_3^{IJ}
\e^J \\
\EuFrak{L}_- \e^I = 0 \\ 
\EuFrak{L}_A \e^I = 0 \\
\EuFrak{L}_i^* \e^I = 0 \\
\EuFrak{L}_m^* \e^I = 0 \\
\EuFrak{L}_{ij} \e^I = \frac{1}{2} \Gamma_{ij} \e^I \\
\EuFrak{L}_{mn} \e^I = \frac{1}{2} \Gamma_{mn} \e^I \\
\EuFrak{L}_{qr} \e^I = \frac{1}{2} \Gamma_{qr} \e^I ~(q,r~\neq~8). 
\end{array}
\eeq

Finally, only the anti commutator of the odd generators remains to be 
computed, and this task can be easily achieved by looking at the nonzero 
terms in the fermionic product ($\chi^I$ and $\psi^J$ are two 
Killing spinors satisfying the same Killing equation):

\begin{equation}
V = \bar{\chi}^I \Gamma^M \psi_2^J \partial_M \nn
\end{equation}

\noindent in fact, it is possible to demonstrate 
\cite{Figueroa-O'Farrill:1999va} that, if the two spinors satisfy Killing 
equation with the same constant $\l$, $V$ must be a linear combination of 
Killing vectors, and the coefficients in front of 
them are the structure constants we are looking for; after some easy algebra, 
we can see that the only non-vanishing component of $V$ is:

\beq
V = - \bar{\chi} \G^- \psi \xi_-  
\eeq

Actually, $V$ can give us only the anti-commutator of the various components
of a single supersymmetry generator: there could be an antisymmetric part
$\sx\{ Q^I_\a,Q^J_\b \dx\} = M_{\a\b} \varepsilon^{IJ}$ with $M_{\a\b}$
and $\varepsilon^{IJ}$ antisymmetric; in this case, $M$ should be a central 
charge, whose (anti-)commutator with all the generators of the algebra
vanish; but since in the parent algebra there was no such generator,
and since no new generator can appear after a Penrose limit, we can conclude
that this antisymmetric part does not appear in our algebra.  
Finally, we can write down the complete set of non-zero (anti-)commutators of 
the superalgebra $\EuFrak{G} = \EuFrak{G}_0 \otimes \EuFrak{G}_1$:

\beq
\begin{array}{c}
\di \sx[ \xi_+,\xi_i \dx] = - \frac{\mu}{\sqrt{2}} \xi_i^* \\
\di \sx[ \xi_+,\xi_m \dx] = - \frac{\mu}{2} \xi_m^* \\
\di \sx[ \xi_+,\xi_i^* \dx] = \frac{\mu}{\sqrt{2}} \xi_i \\
\di \sx[ \xi_+,\xi_m^* \dx] = \frac{\mu}{2} \xi_m \\
\di \sx[ \xi_i^*,\xi_j \dx] = \frac{\mu}{\sqrt{2}} \delta_{ij} \xi_- \\
\di \sx[ \xi_m^*,\xi_n \dx] = \frac{\mu}{2} \delta_{mn} \xi_- \\
\di \sx[ \xi_{AB},\xi_C \dx] = - \eta_{AC} \xi_B + \eta_{BC} \xi_A \\
\di \sx[ \xi_{AB},\xi_C^* \dx] = - \eta_{AC} \xi_B^* + \eta_{BC} \xi_A^* \\
\di \sx[ \xi_{AB},\xi_{CD} \dx] = \eta_{AD} \xi_{BC} + \eta_{BC} \xi_{AD} 
- \eta_{AC} \xi_{BD} - \eta_{BD} \xi_{AC} \\
\di \sx[ \xi_+,Q^I \dx] = \frac{1}{4} \mu \sx(2W+\sqrt{2}Y\dx) \s_3^{IJ}Q^J \\
\di \sx[ \xi_{AB},Q^I \dx] = \frac{1}{2} \G_{AB} Q^I \\
\di \sx\{ Q^I_\a,Q^J_\b \dx\} = - \sx( \G^- C^{-1} \dx)_{\a\b} \d^{IJ} \xi_- 
\end{array}
\eeq

\noindent where $C$ is the charge conjugation matrix and $\xi_{AB}$ is nonzero
 only when $A,B = \{1,2\}$ or $\{3,4\}$ or $\{5,6,7\}$.

\section{Supersymmetric $\sigma$-model Lagrangian}
\label{lagr}

String theory is essentially a 2-dimensional field theory, whose fields $X^M$ 
and $\Theta^I$ are coupled via certain tensors (in IIB: a ``metric'' $g_{MN}$,
 and some ``potentials'' $A_{MN}$, $B_{MN}$ and $F_{M_1\dots M_5}$); in order 
for this theory to be well defined, we need Weyl invariance at the quantum 
level, and then we must impose a set of conditions on the ``metric'' and on 
the ``potentials'': it turns out that those conditions are exactly the 
supergravity field equations.\\
In other words, if we want a consistent string theory Lagrangian, we must
use background fields satisfying supergravity equations of motion.
From now on, we will use upper case X to denote $\sigma$-model fields 
corresponding to target space coordinates. The $\sigma$-model Lagrangian can 
be constructed using the metric $g_{MN}$ and the potential 2-form $B$ defined, 
since the RR potential is set to zero, as:

\beq
\begin{array}{c}
G = d B \\
B_{uM} = G_{uMN} X^N 
\end{array}
\eeq

\noindent where we used the gauge freedom to set $\partial_u B = 0$; in 
components:

\beq
\begin{array}{c}
B_{u i} = 2 \mu \varepsilon_{ij} X^j \\
B_{u m} = \sqrt{2} \mu \varepsilon_{mn} X^n. 
\end{array}
\eeq

The supersymmetric $\sigma$-model Lagrangian can be written according to the 
standard formula \cite{Metsaev:2002re,Cvetic:2002nh}:

\beq
\label{lagrgen}
\begin{array}{c}
\mathcal{L} = \frac{1}{2} \left(\sqrt{-h} h^{ab} g_{MN} - \varepsilon^{ab} 
B_{MN} \right) \partial_a X^M \partial_b X^N + \\
+ i \left(\sqrt{-h} h^{ab} \delta^{IJ} - \varepsilon^{ab} \sigma_3^{IJ} 
\right) \partial_a X^M \bar{\Theta}^I \Gamma_M \mathcal{D}_b^{JK} \Theta^K 
\end{array}
\eeq

\noindent where $h^{ab}$ is the world-sheet metric and $h$ its determinant, 
$\varepsilon$ the Levi Civita tensor and $\sigma_{1,2,3}$ are the usual Pauli 
matrices; $\Theta^I$ (with $I=1,2$) are Majorana-Weyl 10 dimensional spinors, 
whose conjugate is, as usual, $\bar{\Theta}^I = {\Theta^I}^\dagger \Gamma_0$, 
and the derivative operator $\mathcal{D}$ is defined as:

\beq
\begin{array}{c}
\mathcal{D}_a^{IJ} = \partial_a ~\delta^{IJ} + \frac{1}{4} \der_a X^M \sx[ 
\left( \omega_{M~NP} ~\delta^{IJ} - \frac{1}{2} H_{MNP} ~\sigma_3^{IJ} \right)
\Gamma^{NP} + \dx.\\
+ \sx. \sx( \frac{1}{3!} F^{RNP}\Gamma_{RNP} \sigma_1^{IJ} + 
\frac{i}{2\cdot 5!} F^{M_1 \dots M_5}\Gamma_{M_1 \dots M_5}\sigma_2^{IJ} 
\right)\Gamma_M \dx] 
\end{array}
\eeq

\noindent that is, the pull-back of the generalized covariant derivative 
appearing in Killing spinor equation in IIB supergravity, where we divided 
the complex three-form $G$ into its NS and RR parts: $G = H + \Phi F$, where 
$\Phi$ is the dilaton-axion.\\
The Lagrangian written above (\ref{lagrgen}) is invariant under both rigid 
space-time supersymmetry transformations, depending on a constant parameter 
$\eta$:

\beq
\begin{array}{c}
\delta X^M = - i \bar{\Theta}^I \Gamma^M \eta^I \\
\delta \Theta^I = \eta^I 
\end{array}
\eeq

\noindent and local supersymmetry (or $\kappa$-symmetry \cite{Grisaru:1985fv})
 transformations which read \cite{Cvetic:2002nh}:

\beq
\begin{array}{c}
\delta \Theta^I = \left( \iden + \Gamma \right)^{IJ} \kappa^J \\
\delta X^M = i \bar{\Theta}^I \Gamma^M \delta \Theta^I 
\end{array}
\eeq

\noindent where $\kappa$ is a spinor parametrizing this transformation, while 
the matrix $\Gamma$ is defined by:

\begin{equation}
\Gamma^{IJ} = \frac{1}{2\sqrt{- h}} \varepsilon^{ab} \partial_a X^M 
\partial_b X^N \Gamma_{MN} \sigma_3^{IJ}.
\end{equation}

Turning back to the $\sigma$-model, we recall that the only nonzero 
components of the spin connection $\omega$ (\ref{spinconn}) and 3-form field 
$G$ (\ref{newG}) are:

\beq
\begin{array}{c}
\omega_{u~+i} = \frac{1}{2} \mu y_i =\frac{1}{2} \mu X_i \\
\omega_{u~+m} = \frac{1}{4} \mu y_m =\frac{1}{4} \mu X_m \\
G_{uij} = 2\mu \varepsilon_{ij} \\
G_{umn} = \sqrt{2} \mu \varepsilon_{mn}. 
\end{array}
\eeq

Light cone gauge consists of a set of conditions over both bosonic and 
fermionic fields, that is:

\beq
\label{lightcone}
\begin{array}{c}
u = \tau p^+ \sqrt{\ap} \\
h^{ab} = \eta^{ab} \\
\Gamma^+ \Theta^I = 0; 
\end{array}
\eeq

\noindent where $p^+$ is a constant which may be interpreted as center of mass
 momentum of the string along the $u$ direction.\\
We can demonstrate that both pull-back terms give a nonzero contribution only 
along the $\tau$-$u$ direction, when $\partial_a X^M = \partial_\tau u = 
p^+ \sqrt{\ap}$; first of all, let us examine the supercovariant derivative: 
the connection term is nonzero only when $M=u$ and the term $H_{MNP} 
\Gamma^{NP}$ always generates a $\Gamma^+$ when $M\neq u$; with these 
considerations, the general covariant derivative turns out to be:

\begin{equation}
\mathcal{D}^{IJ}_a = \partial_a \delta^{IJ} + \delta^0_a Z \s_3^{IJ}
\end{equation}

\noindent where $Z$ can be expressed in terms of gamma matrices as:

\begin{equation}
Z = - \frac{1}{4} \m p^+ \sqrt{\ap} \sx( 2 \G^{12} + \sqrt{2} \G^{34} \dx).\nn
\end{equation}

\noindent Let us consider instead the term $\bar{\Theta}^I \Gamma_M 
\mathcal{D}^{IJ}_b \Theta^J$: when $X^M = v$ it vanishes because of the light 
cone gauge condition over $\Theta$, while when $M=A$ it vanishes because of 
the symmetry of Gamma matrices combined with the antisymmetry of grassmannian 
variables $\Theta^I$.\\
The final form of the Lagrangian density in light-cone gauge is then:

\beq
\begin{array}{c}
\di \mathcal{L} = \frac{1}{2} \dot{X}^A\dot{X}_A - \frac{1}{2} {X'}^A {X'}_A 
- \frac{1}{4} \sx(\mu p^+ \dx)^2 \ap \sx(X^i X_i + \frac{1}{2} X^m X_m \dx)
 + \\
\di + 2 \mu p^+ \sqrt{\ap} \varepsilon_{ij} X^i {X'}^j + \sqrt{2} \mu p^+ 
\sqrt{\ap} \varepsilon_{mn} X^m {X'}^n +  \\
\di + i p^+ \sqrt{\ap} \bar{\Theta}^1 \G^- \sx( \der_\tau - \der_\s + Z \dx) 
\Theta^1 + i p^+ \sqrt{\ap} \bar{\Theta}^2 \G^- \sx( \der_\tau + \der_\s - 
Z \dx) \Theta^2   
\end{array}
\eeq

\noindent where dot and prime mean respectively derivative with respect to 
$\tau$ and $\sigma$. We can now rescale all the fields in $\mathcal{L}$ 
so that they all are adimensional (in practice, this amounts to sending $\m 
\to \sqrt{\ap} \m$ and considering also $\s$ and $\tau$ adimensional); 
moreover, we reabsorb the factor $p^+ \sqrt{\ap}$ in the definition of 
fermions. 

\section{Spectrum of the Hamiltonian}
\label{hamilton}

The procedure we adopt in order to find the spectrum of our theory, is the 
same in both fermionic and bosonic sectors: since we are dealing with a closed
 string theory, we expand all fields in Fourier 
modes and, by means of the equations of motion, write down the explicit 
dependence of fields on world sheet coordinates; then we promote Fourier 
coefficients to operators and impose canonical (anti)-commutation relations. 
The last step consists of writing explicitly $H$ in terms of 
creation and annihilation operators.

\subsection{Bosonic sector}

Using Euler-Lagrange equations, we deduce the equations of motion for the $X$ 
fields, which are:

\bea
&\ddot{X}^q = {X''}^q \label{qeq}\\
&\ddot{X}^i = - \frac{1}{2} \sx( \mu p^+ \ap \dx)^2 X^i + {X''}^i - 4 
\sx(\mu p^+ \ap \dx) \varepsilon^{ij} {X'}^j \label{ieq}\\
&\ddot{X}^m = - \frac{1}{4} \sx(\mu^2 p^+ \ap \dx)^2 X^m + {X''}^m - 2 
\sqrt{2} \sx( \mu p^+ \ap \dx) \varepsilon^{mn} {X'}^n; \label{meq}
\eea

\noindent since we are dealing with a closed string theory, we must take
into account the periodicity condition on $\sigma$ coordinate ($X(\sigma + 
2 \pi,\tau) = X(\sigma, \tau)$) and we can expand the $X$ fields in Fourier 
modes: 

\begin{equation}
X^A(\sigma,\tau) = \sum_{s:-\infty}^{+\infty} \mathcal{C}_s^A(\tau) e^{is
\sigma} \label{fourier}
\end{equation}

\noindent because of the reality condition that must be imposed on all the $X$ 
fields, Fourier coefficients in (\ref{fourier}) must obey:

\begin{equation}
\mathcal{C}_s^A(\tau) = {\mathcal{C}^A_{-s}}^*(\tau)  
\end{equation}

\noindent now equations (\ref{qeq}), (\ref{ieq}) and (\ref{meq}) can be 
rewritten as a single matricial relation among coefficients, namely:

\begin{equation}
\ddot{\mathcal{C}^A_s} = - T^A_{s~~B} \mathcal{C}_s^B ~~\mathrm{
(no~sum~over~}s) \label{ctc}
\end{equation} 

\noindent where matrix $T_s$ is defined to be ($\l = \m p^+ \ap$):

\begin{equation}
T_s = \begin{pmatrix} \frac{1}{2} \l^2 + s^2 & - 4 i \l s & 0 & 0 & 0 & 0 
& 0 & 0 \\  4 i \l s & \frac{1}{2} \l^2 + s^2 & 0 & 0 & 0 & 0 & 0 & 0 \\ 
0 & 0 & \frac{1}{4} \l^2 + s^2 & - 2 \sqrt{2} i \l s & 0 & 0 & 0 & 0 \\ 0 
& 0 & 2 \sqrt{2} i \l s & \frac{1}{4} \l^2 + s^2 & 0 & 0 & 0 & 0 \\ 0 & 0
 & 0 & 0 & s^2 & 0 & 0 & 0 \\ 0 & 0 & 0 & 0 & 0 & s^2 & 0 & 0 \\ 0 & 0 & 
0 & 0 & 0 & 0 & s^2 & 0 \\ 0 & 0 & 0 & 0 & 0 & 0 & 0 & s^2\end{pmatrix} 
\end{equation}

\noindent whose eigenvalues are:

\beq
\label{eigenvalues} 
\begin{array}{c}
s^2~\mathrm{(4~times)} \\
k^\pm_s = \frac{1}{2} \left( 2 s^2 \pm 8 s \mu p^+ \ap + \sx(\mu p^+ \ap
\dx)^2 \right) \\
j^\pm_s = \frac{1}{4} \left( 4 s^2 \pm 8 \sqrt{2} s \mu p^+ \ap + 
\sx(\mu p^+ \ap \dx)^2 \right)
\end{array}
\eeq

\noindent in order to find a solution to equation (\ref{ctc}) it is convenient 
to introduce a set of orthonormal complex eigenvectors of $T_s$ ($v^\pm$ and 
$w^\pm$, obeying $v^\mp \cdot v^\pm = w^\mp \cdot w^\pm = 1$, all other 
products being 0), corresponding respectively to eigenvalues $k^\pm$ and 
$j^\pm$ in (\ref{eigenvalues}), and reorganize the $X^i$ and $X^m$ fields as 
vectors along $v^\pm$ and $w^\pm$ basis:

\beq
\begin{array}{c}
\di X^{(i)} = \begin{pmatrix} X^1 \\ X^2 \end{pmatrix} = 
\sum_{s=\im}^{\ip} \mathcal{C}_s^{(i)} e^{is\sigma}\\
\di X^{(m)} = \begin{pmatrix} X^3 \\ X^4 \end{pmatrix} = 
\sum_{s=\im}^{\ip} \mathcal{C}_s^{(m)} e^{is\sigma}
\end{array}
\eeq

\noindent where, for every $s$:

\beq
\begin{array}{c}
\mathcal{C}^{(i)}_s = \begin{pmatrix} \mathcal{C}^1 \\ \mathcal{C}^2 
\end{pmatrix} = c^+_s v^+ + c^-_s v^- \\
\mathcal{C}^{(m)}_s = \begin{pmatrix} \mathcal{C}^3 \\ \mathcal{C}^4
\end{pmatrix} = d^+_s w^+ + d^-_s w^- 
\end{array}
\eeq

\noindent now, the reality condition on the $c_s^\pm$ and $d_s^\pm$ read:

\beq
\begin{array}{c}
{c_{-s}^\mp}^* {v^\mp}^* = {c_{-s}^\mp}^* v^\pm = c_s^\pm v^\pm 
~~\Rightarrow ~~{c_{-s}^\mp}^* = c_s^\pm \\
{d_{-s}^\mp}^* {w^\mp}^* = {d_{-s}^\mp}^* w^\pm = d_s^\pm w^\pm 
~~\Rightarrow ~~{d_{-s}^\mp}^* = d_s^\pm 
\end{array}
\eeq

\noindent and the equations of motion, which are greatly simplified with these 
conventions, now read:

\beq
\begin{array}{c}
\ddot{\mathcal{C}}_s^q = - s^2 \mathcal{C}_s^q \\
\ddot{c}_s^\pm = - k_s^\pm c_s^\pm \\
\ddot{d}_s^\pm = - j_s^\pm d_s^\pm 
\end{array}
\eeq

\noindent and can be easily solved in terms of the square roots of $T$ 
eigenvalues $|s|$, $\omega^\pm_s = \sqrt{k^\pm_s}$ and $\nu^\pm_s = \sqrt{j^\pm_s}$:

\beq
\begin{array}{c}
\mathcal{C}_0^q = x^q + p^q \tau \\
\mathcal{C}_s^q = A^q_s e^{- i |s| \tau} + B^q_s e^{i |s| \tau}\\
c^\pm_s = A^\pm_s ~ e^{-i\o_s^\pm \tau} + B^\pm_s ~ e^{i\o_s^\pm \tau}\\
d^\pm_s = C^\pm_s ~ e^{-i\n_s^\pm \tau} + D^\pm_s ~ e^{i\n_s^\pm \tau}. 
\end{array}
\eeq

Reality condition can be written in terms of the new coefficients, and become:

\beq
\begin{array}{c}
{B_{-s}^q}^* = A_s^q \\
{B_{-s}^\mp}^* = A_s^\pm \\
{D_{-s}^\mp}^* = C_s^\pm  
\end{array}
\eeq

\noindent we can write down $X$ fields as:

\beq
\begin{array}{c}
\di X^q = x^q + p^q \tau + \sum_{s \neq 0} \sx( A^q_s e^{- i |s| \tau} + 
{A^q_{-s}}^* e^{i |s| \tau}\dx) e^{i s \s} \\
\di X^{(i)} = \sum_{s=\im}^{\ip} \sx[ \sx(A_s^+ e^{- i \o_s^+ \tau} + 
{A_{-s}^-}^* e^{i \o_s^+ \tau}\dx) v^+ + \sx(A_s^- e^{- i \o_s^- \tau} + 
{A_{-s}^+}^* e^{i \o_s^- \tau}\dx) v^- \dx] e^{i s \s} \\
\di X^{(m)} = \sum_{s=\im}^{\ip} \sx[ \sx(C_s^+ e^{- i \n_s^+ \tau} + 
{C_{-s}^-}^* e^{i \n_s^+ \tau}\dx) w^+ + \sx(C_s^- e^{- i \n_s^- \tau} + 
{C_{-s}^+}^* e^{i \n_s^- \tau}\dx) w^- \dx] e^{i s \s} 
\end{array}
\eeq

\noindent equipped with these definitions, we can write the momenta 
conjugate to target space coordinates $X$:

\beq
\Pi_A = \frac{\delta L}{\delta \dot{X}^A} = \dot{X}_A \nn
\eeq

\noindent which explicitly read:

\beq
\begin{array}{c}
\di \Pi^q = p^q - \sum_{s \neq 0} i |s| \sx( A^q_s e^{-i |s| \tau} - 
{A_{-s}^q}^* e^{i |s| \tau} \dx) e^{i s \s} \\
\di \Pi^{(i)} = \sum_{s=\im}^{\ip} -i \sx[ \o_s^\pm \sx( A^\pm_s  
e^{-i \o_s^\pm \tau} - {A^\mp_{-s}}^* e^{i \o_s^\pm \tau}\dx) v^\pm 
\dx] e^{i s \s} \\
\di \Pi^{(m)} = \sum_{s=\im}^{\ip} -i \sx[\nu_s^\pm \sx( C^\pm_s 
e^{-i \nu_s^\pm \tau} - {C^\mp_{-s}}^* e^{i \nu_s^\pm \tau}\dx) w^\pm
\dx] e^{i s \s} 
\end{array}
\eeq

\noindent it is possible to invert the above relations and get an expression 
for the Fourier coefficients: 

\beq
\begin{array}{c}
\di x^q = \int_0^{2\pi} \frac{d\s}{2\pi} \sx(X^q - \tau \Pi^q \dx)\\
\di p^q = \int_0^{2\pi} \frac{d\s}{2\pi} \Pi^q \\
\di A_s^q = \frac{1}{2} \int_0^{2\pi} \frac{d \s}{2 \pi} \sx( X^q
 + \frac{i}{|s|} \Pi^q \dx) e^{i (|s| \tau - s\s)} \\
\di A_s^\pm = \frac{1}{2} \int_0^{2 \pi} \frac{d \s}{2 \pi} \sx( 
X^{(i)} + \frac{i}{\o_s^\pm} \Pi^{(i)}\dx) \cdot v^\mp e^{i(\o_s^\pm \tau 
- s \s)} \\
\di C_s^\pm = \frac{1}{2} \int_0^{2 \pi} \frac{d \s}{2 \pi} \sx( 
X^{(m)} + \frac{i}{\nu_s^\pm} \Pi^{(m)}\dx) \cdot w^\mp e^{i(\nu_s^\pm \tau 
- s \s)}
\end{array}
\eeq

\noindent in order to quantize this Lagrangian, we promote $X$ and $\Pi$ to 
operators, imposing canonical commutation relations:

\beq
\label{commutation}
\left[ X^A(\tau, \sigma), \Pi_B(\tau, \sigma') \right] = i \delta^A_B \delta(
\sigma - \sigma'). 
\eeq

Condition (\ref{commutation}) imply that, when we interpret Fourier 
coefficients as operators ($A^* \rightarrow A^\dagger$):

\beq
\begin{array}{c}
\left[ x^q, p^{q'} \right] = \frac{i}{4 \pi} \eta^{qq'} \\
\sx[ A_s^q , {A_{s'}^{q'}}^\dagger \dx] = \frac{1}{8 \pi |s|} \d_{s s'} 
\eta^{qq'} \\
\sx[ A_s^\pm , {A_{s'}^\pm}^\dagger \dx] = \frac{1}{8 \pi \o_s^\pm} 
\d_{s s'} \\
\sx[ C_s^\pm , {C_{s'}^\pm}^\dagger \dx] = \frac{1}{8 \pi \n_s^\pm} 
\d_{s s'} 
\end{array} 
\eeq

\noindent all the above operators can be rescaled in order to have canonical 
commutation relations for creation and annihilation operators:

\beq
\begin{array}{c}
\hat{x}^q~,~\hat{p}^q = 2 \sqrt{\pi}~ x^q~,~p^q \\
a_s^q = 2 \sqrt{2 \pi |s|} A_s^q \\
a_s^\pm = 2 \sqrt{2 \pi \o_s^\pm} A_s^\pm \\
c_s^\pm = 2 \sqrt{2 \pi \nu_s^\pm} C_s^\pm 
\end{array}
\eeq

Now we can construct the Hamiltonian, with a Legendre transformation:

\begin{equation}
H_{bos} = \int d\s \sx( \Pi_A X^A - \mathcal{L}_{bos} \dx) 
\end{equation}

\noindent after some algebra, we can write an explicit expression for 
$H$ in terms of $X$ and $\Pi$:

\beq
\begin{array}{c}
H_{bos} = H_{bos}^{(q)} + H_{bos}^{(i)} + H_{bos}^{(m)} \\
H_{bos}^{(q)} = \frac{1}{2} \dot{X}^q \dot{X}_q + \frac{1}{2} 
{X'}^q {X'}_q \\
H_{bos}^{(i)} = \frac{1}{2} \dot{X}^i \dot{X}_i + \frac{1}{2} 
{X'}^i {X'}_i + \frac{1}{4} \sx(\mu p^+ \ap \dx)^2 X^i X_i - 2 \mu p^+ \ap 
\ve_{ij} X^i {X'}^j \\
H_{bos}^{(m)} = \frac{1}{2} \dot{X}^m \dot{X}_m + \frac{1}{2} 
{X'}^m {X'}_m + \frac{1}{8} \sx(\mu p^+ \ap \dx)^2 X^m X_m - 2 \sqrt{2} 
\mu p^+ \ap \varepsilon_{mn} X^m {X'}^n 
\end{array}
\eeq

\noindent using now the definition of $X$ in terms of creation and 
annihilation operators, we can write $H$ in terms of number 
operators, namely:

\beq
\label{hambose}
\begin{array}{c}
  \di H_{bos}^{(q)} = \frac{1}{4} \hat{p}^q \hat{p}_q + \sum_{s=\im}^{\ip} 
\frac{|s|}{4} \sum_q \sx( 2\mathcal{N}_s^{q} + 1 \dx) \\
  \di H_{bos}^{(i)} = \sum_{s=\im}^{\ip} \sx[ \o_s^\pm \sx( 2
\mathcal{N}_s^{(i)~\pm} + 1 \dx) \dx] \\
  \di H_{bos}^{(m)} = \sum_{s=\im}^{\ip} \sx[ \nu_s^\pm 
\sx( 2 \mathcal{N}_s^{(m)~\pm} + 1 \dx) \dx] 
\end{array}
\eeq

\noindent where

\beq
\begin{array}{c}
  \mathcal{N}_s^q = a^{q~\dagger}_s a_{s~q} \\
  \mathcal{N}_s^{(i)~\pm} = - a_s^{(i)~\pm~\dagger} \cdot a_s^{(i)~\pm}\\
  \mathcal{N}_s^{(m)~\pm} = - c_s^{(m)~\pm~\dagger} \cdot c_s^{(m)~\pm}
\end{array}
\eeq

\noindent as it was expected, the Hamiltonian is proportional 
to number operators. There is one more condition on these states, which can be
 obtained imposing that a translation in $\s$ does not affect the physical 
result; the operator generating $\s$-translations is:

\begin{equation}
P = \int d\s \Pi^A \der_\s X_A.
\end{equation}

After some algebra on creation and annihilation operators, we obtain that 
physical states must obey:

\beq
\begin{array}{c}
  \mathcal{N}^{q~R} = \mathcal{N}^{q~L} \\
  \mathcal{N}^{(i)~+} = \mathcal{N}^{(i)~-} \\
  \mathcal{N}^{(m)~+} = \mathcal{N}^{(m)~-} 
\end{array}
\eeq

\noi where $\mathcal{N}^{q~R,L}$ represents the total number of left ($n > 0$) 
or right movers ($n<0$) in the flat sector, while $\mathcal{N}^{(i),(m)~\pm} 
= \sum_s \mathcal{N}^{(i),(m)~\pm}_s$ represents the total number of $(i)$ \
or $(m)$ ``left'' ($+$) or ``right'' ($-$) modes: it is again a sort of {\it 
level-matching} condition.

\subsection{Fermionic sector}

We start with the fermionic equations of motion, which read:

\beq
\label{fermieom} 
\begin{array}{c}
\sx( \der_\tau -\der_\s + Z \dx) \Theta^1 = 0 \\
\sx( \der_\tau +\der_\s - Z \dx) \Theta^2 = 0 
\end{array} 
\eeq

\noindent In order to solve these equations, we expand the fermionic fields 
$\Theta^I$ in Fourier modes, taking into account periodicity condition over 
$\sigma$ and reality of both spinors:

\beq
\begin{array}{c}
\di \Theta^I = \sum_{n:\im}^{\ip} \Theta_n^I( \tau ) e^{in\s} \\
\di {\Theta_{-n}^I}^* = \Theta_n^I 
\end{array}
\eeq

\noindent it is important to notice that, because of the two conditions (light
 cone and Weyl) we imposed on $\Theta$, these spinor fields have only 8 
nonzero components (and we may choose them to be the upper ones), such that 
(see appendix \ref{conventions}):

\begin{equation}
\Theta^I = \begin{pmatrix} S^I \\ 0 \\ 0 \\ 0 \end{pmatrix}~,~\Theta_n^I = 
\begin{pmatrix} S_n^I \\ 0 \\ 0 \\ 0 \end{pmatrix} 
\end{equation}

\noindent Since the matrix $Z$ has a block-diagonal form, we can simplify 
equations of motion as (see appendix \ref{conventions} for conventions 
on gamma matrices):

\beq
\left[ \partial_\tau \delta^{IJ} - \left( \partial_\sigma + \frac{1}{4} 
\sx(\mu p^+ \ap \dx) \sx( \sqrt{2} c^{34} + 2 c^{12} \dx) \right) 
\sigma_3^{IJ} \right] S^I = 0
\eeq

\noindent like in the bosonic case, we introduce a set of orthonormal 
eigenvectors $u_\alpha$ ($\alpha$ runs over spinorial indices $1 \dots 8$, and $u_\a^* \cdot u_\b = \delta_{\a\b}$) 
of the matrix $M_n = n \iden + i Z$ , such that:

\beq
\begin{array}{c}
\di S_n^I = \sum_{\a=1}^8 \zeta^I_{n~\a}(\tau) u_\a \\
M_n u_\a = \o_{n~\a} u_\a ~\mathrm{(no~sum~on}~\a) 
\end{array}
\eeq

\noindent where the $\o_{n~\a}$ are:

\beq
\begin{array}{c}
\o_{n~1} = \o_{n~2} = \sx[ n - \frac{1}{4} \m p^+ \ap (2 + \sqrt{2}) \dx] \\
\o_{n~3} = \o_{n~4} = \sx[ n + \frac{1}{4} \m p^+ \ap (2 - \sqrt{2}) \dx] \\
\o_{n~5} = \o_{n~6} = \sx[ n + \frac{1}{4} \m p^+ \ap (\sqrt{2} - 2) \dx] \\
\o_{n~7} = \o_{n~8} = \sx[ n + \frac{1}{4} \m p^+ \ap (2 + \sqrt{2}) \dx] 
\end{array}
\eeq

\noindent and the following relations among eigenvectors and eigenvalues hold:

\beq
\begin{array}{c}
{u_\a}^* = u_{(9-\a)} \\
\o_{-n~ \sx( 9 - \a \dx) } = - \o_{n~\a} 
\end{array}
\eeq

\noindent so that the reality condition becomes:

\begin{equation}
\sx(\zeta_{-n~\sx(9-\a\dx)}^I\dx)^* = \zeta^I_{n~\a}
\end{equation}

\noi now the equations of motion read:

\beq
\dot{\zeta}^I_{n~\a} = i \s_3^{IJ} \o_n^\a \zeta^J_{n~\a}~
\mathrm{(no~sum~over~\a)} 
\eeq

\noindent which can be easily integrated:

\begin{equation}
\zeta^I_{n~\a} = \psi^I_{n~\a} e^{i \s_3^{II} \o_n^\a \tau}. 
\end{equation}

Finally, spinors $S^I$ can be written as:

\begin{equation}
\displaystyle S^I = \sum_{n=\im}^{\ip} \sum_{\a=1}^{8} \psi_{n~\a}^I e^{i 
\s_3^{II} \o_n^\a \tau} e^{i n \s} u_\a
\end{equation}

\noindent this relation can be inverted to obtain Fourier coefficients:

\begin{equation}
\psi^I_{m~\b} = \int \frac{d\s}{2\pi} e^{- i m \s} u_\b^* \cdot S^I 
e^{- i \s_3^{II} \o_m^\b \tau} 
\end{equation}

\noindent as usual in IIB string theory $\sigma$-model, the two spinors 
correspond to right and left movers  along the closed string, depending on
the sign of $m$; we can now find the momenta conjugate to fermionic 
coordinates, which are:

\begin{equation}
\pi^I = \frac{\d \mathcal{L}}{\d \der_\tau S^I} = - i {S^I}^* 
\G^-~~~~\bar{\pi}^I = \frac{\d \mathcal{L}}{\d \der_\tau \bar{S}^I} = 0 \nn
\end{equation}

\noindent using these definitions, we can construct the fermionic part of the 
Hamiltonian $H_{fermi}$ which turns out to be:

\beq
\begin{array}{c}
\di H_{fermi} = \int d\s \sx( \pi^I \der_\tau S^I + \der_\tau {S^I}^* 
\bar{\pi}^I - \mathcal{L}_{fermi} \dx) = \\
\di = i \int d\s \sx[ {S^I}^* \G^- \sx( \der_\s - Z \dx)
\s_3^{IJ} S^J \dx]
\end{array}
\eeq

\noindent with the help of the equations of motion (\ref{fermieom}), we can 
simplify the above expression, getting:

\begin{equation}
H_{fermi} = i \int d\s ~{S^I}^* \G^- \der_\tau S^I. 
\end{equation}

In order to quantize this Hamiltonian, we impose canonical anticommutation 
relations on fermionic fields, which read:

\beq
\begin{array}{c}
\sx\{ S_\a^I (\s,\tau), {S^J_\b}^\dagger (\s ', \tau)\dx\} = \d^{IJ} \d_{\a 
\b} \d (\s - \s ') \\
S^I_\a = u_\a^* \cdot S^I 
\end{array}
\eeq

\noindent these conditions imply analogue ones over operators $\psi$ 
and $\psi^*$:

\begin{equation}
\sx\{ \psi_{n~\a}^I , {\psi_{m~\b}^J}^* \dx\} = \frac{1}{2\pi} 
\delta_{mn} \delta_{\a \b} \delta^{IJ} 
\end{equation}

\noindent we can define creation and annihilation operators, by means of the 
following relations:

\beq
\begin{array}{c}
\xi_{m~\alpha}^I = \sqrt{2\pi} {\psi_{m~\alpha}^I}^* \\
{\xi_{m~\alpha}^I}^\dagger = \sqrt{2\pi} \psi_{m~\alpha}^I 
\end{array}
\eeq

\noindent correspondingly, the Hamiltonian becomes:

\begin{equation}
\di H_{fermi} = \sum_\a \sx[ \sum_I \sum_{n=\im}^{\ip} 
\o_{n~\a} \sx( 2 \mathcal{N}^I_{n~\a} - 1 \dx) \dx] \label{hamfermi}
\end{equation}

\noindent where as usual:

\beq
\mathcal{N}^I_{n~\alpha} = {\xi_{n~\alpha}^I}^\dagger \xi_{n~\alpha}^I. 
\eeq

It is worthwhile to stress again the fact that $\sigma$-model coming from 
pp-wave type solutions to IIB supergravity is solvable, being quadratic 
in both fermionic and bosonic fields in light-cone gauge. Another interesting 
remark is in order: despite the solution has supersymmetries, the masses of 
bosonic and fermionic particles are different, and hence the string has a non 
null zero-point energy (analogous features in different backgrounds were seen 
in \cite{Casero:2002ks,Itzhaki:2002kh}), as we can see from the structure of 
the eigenvalues of the Hamiltonian: 

\bea
&\o_n^{bos} = \sqrt{n^2 + a n \sx(\m p^+ \ap\dx) + b \sx(\m p^+ \ap \dx)^2} 
\nn \\
&\o_n^{fer} = n + \frac{1}{4} c \sx(\m p^+ \ap \dx) \nn
\eea

where the coefficients $a$, $b$ and $c$ take the values for the fields $X^A$ 
and the components of $\Theta^I$:

\bea
&a = \sx(0,0,0,0,4,4,2\sqrt{2},2\sqrt{2} \dx) \nn\\
&b = \sx(0,0,0,0,\frac{1}{2},\frac{1}{2},\frac{1}{4},\frac{1}{4} \dx) \nn\\
&c = \sx( -2-\sqrt{2},-2-\sqrt{2},2-\sqrt{2},2-\sqrt{2},-2+\sqrt{2},
-2+\sqrt{2},2+\sqrt{2},2+\sqrt{2} \dx). \nn
\eea

The reason for this asymmetry is easy to explain: when we impose fermionic 
gauge 
conditions, $(\Gamma^+ \Theta = 0)$ we are fixing a certain amount of 
supersymmetry; the residual supersymmetry transformations should then
preserve not only the equations of motion (and this is guaranteed by the
explicit supersymmetric construction of the Lagrangian), but also the gauge
fixing conditions. It can be demonstrated (see \cite{Cvetic:2002nh}) that 
half of the components of the parameter $\k^I$ should be used as compensator
of $\e^I$ variation in order to maintain condition $\G^+ \Theta= 0$; then it
turns out that light-cone condition $X^+ = p^+ \sqrt{\ap} \tau$ is always 
satisfied; the residual supersymmetry transformations preserving both
the Lagrangian and the gauge conditions can be divided into two sets: there 
are shift (inhomogeneous) supersymmetries, involving only the fermionic 
fields and depending on Killing vectors $\e^I$ such that $\G^+ \e^I =0$, and 
homogeneous supersymmetries, that have the usual form and depend on parameters
$\e^I$ such that $\G^- \e^I =0$: the so called ``supernumerary'' Killing
spinors. Notice that this asymmetry in the masses is only due to the gauge
fixing: we could have chosen, for instance, the so-called physical gauge 
($X^0 = \tau$, $X^9 = \sigma$ and $\Gamma_{1 \dots 8} \Theta = 0$) 
\cite{Cvetic:2002nh}, where the Hamiltonian is no more easily quantizable, 
but the supersymmetries are explicit in the model.

\section{BMN Duality}
\label{maldacena}

BMN duality consists of a particular limit on both sides
of $AdS/CFT$ correspondence: on the gravity side, it corresponds to a Penrose 
limit, which permits us to get a quantizable background of IIB string theory; 
on the $CFT$ side, it corresponds to a large $J$ and $\D$ limit, with finite
$\D - J$: the precise statement by BMN is that there exists a certain set of 
non-BPS operators whose two-point function reproduces exactly the string 
spectrum. 
In order to test this conjecture, the first step consists in identifying the 
$CFT$ living on the border of $AdS$; the $AdS_3 \otimes S_3 \otimes S_3 
\otimes S_1$ configuration can be obtained as a near horizon limit from a 
double system of $D1-D5$ branes characterized by:

\beq 
N_5^{(1)} = N_5^{(2)}=N_5
\eeq

where $N_a^{(i)}$ is the number of $Da$branes of the $i$ system; the dilaton
is set to a constant

\beq
e^{- \phi} = \sqrt{2}
\eeq

and $N_5$ is related to the number of parallel coincident $D_1$-branes:

\beq
N_5 = N_1 = \sqrt{2 N_1^{(1)} N_1{(^2)}}
\eeq

moreover, the radii of the 3-spheres are proportional to the number of 
$D1$-branes:

\beq
\begin{array}{c}
  R_{AdS} = \sqrt{\frac{1}{2} \ap N_1} \\
  R_{Sph} = \sqrt{\ap N_1}  
\end{array}
\eeq

while the radius of the $S^1$ is not fixed by the configuration; as shown in 
\cite{deBoer:1999rh}, the symmetry algebra of the $AdS$ background is 

\begin{equation}
SO(2,2) \otimes SO(4) \otimes SO(4) \otimes U(1)  
\end{equation}

\noi which can be rewritten as:

\begin{equation}
\sx( Sl(2,R) \otimes SU(2) \otimes SU(2) \dx)^2 \otimes U(1) \label{alg}
\end{equation}

The dual $CFT$ can be shown  \cite{Giveon:2003ku} to be one of the
$\mathcal{N} = (4,4)$ double SCFT based on the $A_\g$ algebras, found 
in \cite{Sevrin:1988ew}, which are the Kac-Moody extension of (\ref{alg}).
$A_\g$ algebras are characterized by two independent parameters: either the 
levels of the two affine $SU(2)$'s or the central charge and $\g$, which
measures the relative weight of the levels $k^\pm$; since the levels $k^\pm$ 
are related to the radii of the two 3-spheres, we can conclude that $k^+ = 
k^-$. Actually, the levels of the affine algebras $\hat{SU(2)}$ are:

\begin{equation}
k^\pm = \frac{R^2_{Sph}}{4 R_{AdS} G^{(3)}_N} = \frac{4 R^7 L}{\pi g^2 \ap^4}
\end{equation}

\noi and are related to the central charge $c$ and to $\g$ by the relations:

\beq
\begin{array}{c}
\di c = \frac{6 k^+ k^-}{k^+ + k^-} \\
\di k = k^+ + k^- = \frac{c}{6 \g \sx( 1 - \g \dx)}
\end{array}
\eeq

The central charge is instead given by the formula:

\beq
c= \frac{3 R_{AdS}}{2 G^{(3)}_N}\nn
\eeq  

\noi where $G^{(3)}_N$ is the three-dimensional Newton constant, which can be 
obtained as:

\begin{equation}
G^{(3)}_N = \frac{G^{(10)}_N}{\mathrm{Vol}(S_3 \otimes S_3 \otimes S_1)} \nn
\end{equation}

\noi with  

\begin{equation}
\mathrm{Vol}(S^3) = 2 \pi^2 R_{Sph}^3~~,~~\mathrm{Vol}(S^1) = 2 \pi L \nn
\end{equation}

and $L$ is the radius of $S^1$; we can conclude that

\begin{equation}
G^{(10)}_N = 8 \pi^6 g^2 {\a'}^4 ~~\rightarrow~~ G^{(3)}_N = \frac{\pi g^2 
\ap^4}{8 R^6 L} 
\end{equation}    

and then:

\begin{equation}
c=12 \frac{R^7 L}{\pi g^2 \ap^4}
\end{equation}

from which we can conclude that $\g = \frac{1}{2}$.\\
The $A_\g$ algebras are defined via the OPE's of their generators
\cite{Sevrin:1988ps}:

\beq
\begin{array}{c}
\di T(z) T(w) = \frac{c}{2(z-w)^4} + 2 \frac{T(w)}{(z-w)^2} + 
\frac{\der T(w)}{(z-w)} + \cdots \\
\di T(z) \phi(w) = \frac{h \phi(w)}{(z-w)^2} + \frac{\der \phi(w)}{(z-w)} 
+ \cdots \\
\di G_a(z) G_b(w) = \frac{2c\delta_{ab}}{3(z-w)^3} + \frac{2M_{ab}(w)}{
(z-w)^2} + \frac{\sx(2T(w) \delta_{ab} + \der M_{ab}(w)\dx)}{(z-w)} 
+ \cdots \\
\di A^{\pm i}(z) G_a(w) = \a^{\pm i ~~ b}_{~~~a} \frac{G_b(w)}{(z-w)} \mp 
\frac{2 k^\pm Q_b (w)}{k(z-w)^2} + \cdots \\
\di A^{\pm i}(z)A^{\pm j}(w) = \frac{\epsilon^{ijk} A^{\pm}_k(w)}{(z-w)} 
-\frac{k^\pm \delta^{ij}}{2(z-w)^2} + \cdots \\
\di Q_a(z) G_b(w) = \frac{2\sx( \a^{+ i}_{ab} A^+_i(w) - \a^{- i}_{ab} 
 A^-_i(w) \dx)}{(z-w)} + \frac{\delta_{ab} U(w)}{(z-w)} + \cdots \\
\di A^{\pm i}(z)Q_a(w) = \a^{\pm i~~ b}_{~~~a} \frac{Q_b(w)}{(z-w)}
+\cdots \\
\di U(z) G_a(w) = \frac{Q_a(w)}{(z-w)^2} +\cdots \\
\di Q_a(z) Q_b(w) = -\frac{k\delta_{ab}}{2(z-w)} +\cdots \\
\di U(z) U(w) = -\frac{k}{2(z-w)^2} +\cdots 
\end{array}
\eeq

where $\phi$ stands for all the fields $G$, $A$, $U$ and $Q$ which have 
conformal weight $h$ equal to $\frac{3}{2}$, $1$, $1$ and $\frac{1}{2}$ 
respectively; in complex notations $i=\{+,-,3\}$, $a=\{+,-,+K,-K\}$ and the
non-vanishing values (up to symmetry) of the various symbols are

\bea
& \di M_{ab} \equiv -\frac{4}{k} \sx[ k^- \a^{+i}_{ab} A^+_i + k^+\a^{-i}_{ab} 
A^-_i \dx] \nn \\
& \di \d_{+-} = \d_{+K\, -K}=\frac{1}{2} \quad\qquad 
\e^{+-}_{~~3}=-2i\qquad\qquad\quad \e^{3\pm}_{~~\pm}= \mp i \nn \\
& \di \a^{\pm 3}_{+ -} = -\frac{i}{4} \qquad\qquad\qquad 
\a^{\pm 3}_{+K\, -K} = \mp\frac{i}{4} \qquad\qquad 
\a^{+ -}_{+\, +K} = \frac{i}{2} \nn \\
& \di \a^{++}_{-\, -K} = -\frac{i}{2} \qquad\qquad\qquad 
\a^{- +}_{-\, +K} = - \frac{i}{2} \qquad\qquad 
\a^{- -}_{+\, -K} = \frac{i}{2} \nn
\eea

The explicit study of BMN conjecture in this framework is postponed 
to a future work, but a brief algebraic analysis can help us in guessing 
some feature of BMN duality in this case: we need to identify states which 
have large R-charge J and conformal dimension $\Delta$.
In the case at hand, J can be seen as a $U(1)$ generator in the 
$SU(2)^+_1 \oplus SU(2)^+_2$ algebra related to first of the 3-spheres: 
from this point of view, the bosonic fields that are ``rotated'' by the 
first $SO(4)$ have a well defined non-zero $R$-charge $J$, and correspond 
respectively to $Z$ and $\phi$ fields in (\cite{Berenstein:2002jq}) while 
the other four fields have by definition $J=1$, since they belong to the 
second $SO(4)$ and hence are not rotated so that they are not in 
relation with BMN operators in $CFT$, that is, they do not receive 
corrections to conformal dimension; this is in perfect agreement with string 
theory spectrum, since in $4$ directions we have flat space spectrum $\sx(
\D - J \dx)_s = s$, while in the others we have the expected deviations. 

\section{Conclusions}

In summary, we found the Penrose limit of a particular background of IIB 
supergravity (\ref{dsquadro},\ref{Gform}), and studied its behaviour, which 
has proven to be the expected one, being a Cahen-Wallach space 
\cite{Blau:2002mw}. Then we showed that in this background there is no 
supersymmetry enhancement, since only one half of the supersymmetry charges 
are conserved as in the parent background \cite{Castellani:2000nb}. 
We also found the generators of the superalgebra of the pp-wave solution 
and their commutators and showed that the algebra can be written as 
$\EuFrak{h}(4) \rtimes \left( \EuFrak{so}(2) \oplus \EuFrak{so}(2) \right) 
\oplus \EuFrak{cso}(4) \oplus \EuFrak{so}(2)$, where $\EuFrak{h}$ is a 
Heisenberg algebra while $\EuFrak{cso}$ stands for a contracted 
$\EuFrak{so}$ algebra. We built the supersymmetric $\sigma$-model 
Lagrangian and quantized the corresponding Hamiltonian showing that it is 
solvable, being quadratic in both fermionic and bosonic fields; moreover, 
we found that in this model, being the number of supersymmetry charges the 
minimum for such a solution, i.e. there are no ``supernumerary'' Killing 
spinors, the masses of fermionic fields differ from those of the bosonic 
ones \cite{Cvetic:2002nh}, and this leads to a non null zero-point energy.
In the last chapter, a brief comment on BMN duality is given, discussing 
the CFT algebra in relation to string theory spectrum; a complete and 
explicit analysis remains still to be performed, but on algebraic grounds 
we showed that some expected behaviour holds.

\section*{Acknowledgements}

I would like to thank L. Castellani for very useful discussions and 
enlightening comments on a preliminary version of the paper. 
Work supported in part by the European Community's Human Potential 
Program under contract HPRN-CT-2000-00131 Quantum Space-Time, in which 
L. S. is associated to Torino University

\appendix

\section{Conventions and Gamma Matrices}
\label{conventions}

We summarize the conventions adopted in this paper:

\begin{itemize}
\item a,b (world sheet indices corresponding to $\sigma$ and $\tau$) run over
 0,1
\item M,N run from 0 to 9
\item A,B from 1 to 8
\item i,j from 1 to 2 (labeling the coordinates trivially involved in the 
limit in $AdS$)
\item m,n from 3 to 4 (labeling the coordinates trivially involved in the 
limit in the first $S^3$)
\item q,r from 5 to 8 (labeling the coordinates in the last $S^3$ and $S^1$ 
after the limit)
\item I,J from 1 to 2 (labeling fermionic fields $\Theta$)
\item $\alpha$, $\beta$ from 1 to 8 (spinorial indices)
\end{itemize}

\noindent The world-sheet metric $\eta_{ab}$ is:

\beq
\eta_{00}=-\eta_{11}=1 \nn
\eeq

\noindent while tangent space metric is taken to be ``mostly minus'':

\beq
\label{flatmetric}
\begin{array}{c}
  \eta_{++}= ~\eta_{--} = 0 \\
  \eta_{+-}= ~\eta_{-+} = 1 \\
  \eta_{AB}= - \delta_{AB}
\end{array}
\eeq

\noindent the Levi-Civita tensor is:

\beq
\varepsilon^{01}= -\varepsilon^{10} = 1
\eeq

\noindent and the $\sigma$ are the usual Pauli matrices; for the gamma 
matrices, we adopt a convention similar to that in \cite{Metsaev:2001bj}, 
that is:

\beq
\Gamma_0 = \begin{pmatrix} 0&\gamma_9 \\ \gamma_9&0 \end{pmatrix}~,~ 
\Gamma_A = \begin{pmatrix} 0&\gamma_A \\ \gamma_A&0 \end{pmatrix}~,~
\Gamma_9 = \begin{pmatrix} 0&\iden \\ -\iden&0 \end{pmatrix} 
\eeq

\noindent where $\gamma_A$ are SO(8) gamma matrices obeying:

\beq
\{ \gamma_A,\gamma_B \} = - 2 \delta_{AB}\nn
\eeq

\noindent while $\gamma_9$ is defined as:

\beq
\gamma_9 =- \gamma_1 \dots \gamma_8 ~~,~~ \gamma_9^2 = \iden. \nn
\eeq

The matrices $\gamma_A$ are analogously written by means of $SO(7)$ 
gamma matrices ${\bf{C}}_A$:

\beq
\begin{array}{c}
\di \g_A = \begin{pmatrix} 0&{\bf{C}}_A \\ \tilde{\bf{C}}_A&0 \end{pmatrix}\\
\di \tilde{\bf{C}}_A = {\bf{C}}_A~(A=1 \dots 7), -{\bf{C}}_A~(A=8)\\
\di {\bf{C}}_8 = \iden 
\end{array}
\eeq

\noindent where ${\bf{C}}_A$ matrices are the octonion structure constants:

\bea
&{\bf{C}}_A = \left( {\bf{C}}_A \right)_{BC} \nn \\
&{\bf{C}}_{123}={\bf{C}}_{165}={\bf{C}}_{257}={\bf{C}}_{354}={\bf{C}}_{367}={
\bf{C}}_{246}={\bf{C}}_{147}=1 \\
&{\bf{C}}_{AB8}=\delta_{AB} \nn
\eea

\noindent and antisymmetrization over the 3 indices is intended. \\
The matrix $\Gamma_{11}$ is:

\beq
\Gamma_{11} =- \Gamma_0 \dots \Gamma_9 = \begin{pmatrix} -\iden&0\\0&\iden
\end{pmatrix} 
\eeq

\noindent moreover, due to our choice for the metric (\ref{flatmetric}) we 
can say that:

\beq
\Gamma^+ \Gamma^+ = \Gamma^- \Gamma^- = 0.
\eeq

Matrices $W$ and $Y$ are:

\beq
W = \Gamma^1 \Gamma^2~~,~~ Y = \Gamma^3 \Gamma^4 \nn
\eeq

\noindent and because of their definition, satisfy some important properties:

\beq
\begin{array}{c}
\di W^2 = Y^2 = -\iden~~,~~ \left[W,Y\right] = 0  \\
\di \left[W,\Gamma^\pm \right] = \left[Y,\Gamma^\pm \right] = 0 
\end{array}
\eeq

\noindent notice that, due to the above properties, we are able to integrate 
the gravitino supersymmetry equation (\ref{gravitino}) and to write the 
solution in the form of equation (\ref{residualsusy}).
Because of Weyl condition ($\Gamma_{11} \Theta^I = -\Theta^I$, as the
 gravitino), $\Theta^I$ has only 16 nonzero components, and we can 
write:

\beq
\Theta^I = \begin{pmatrix} \theta^I \\ 0 \end{pmatrix}. 
\eeq

Due to light-cone gauge, we have an additional condition over $\Theta$:

\beq
\Gamma^+ \Theta^I = 0 \rightarrow \sx( \iden + \gamma^9 \dx) \theta^I = 0
\eeq

\noindent this condition halves again the nonzero components of $\Theta^I$, 
but in our basis:

\beq
\gamma^9 = \begin{pmatrix}- \iden_8&0 \\ 0&\iden_8 \end{pmatrix} \nn
\eeq

\noindent so, the only 8 components of $\Theta^I$ surviving are:

\beq
\Theta^I = \begin{pmatrix} S^I_8 \\ 0_8 \\ 0_8 \\ 0_8 \end{pmatrix}. 
\eeq  

\section{$AdS_n$ and $S_n$ spaces}
\label{adsandsphere}

The spaces of the type $AdS_n$ or $S_n$ can be viewed as hyper-surfaces in 
$\mathbb{R}^{n+1}$; in particular, let $z_\mu$ ($\mu = 1 \dots n+1$) be the 
coordinates of $\mathbb{R}^{n+1}$, then $AdS_n$ space is the hyperboloid:

\beq
\di z_0^2 - \sum_{i=1}^{n-1} z_i^2 + z_n^2 = - R^2  
\eeq

\noindent defined in $\mathbb{R}^{n+1}$ with metric $(+,-, \dots ,-,+)$, 
while the n-sphere is: 

\beq
\di \sum_{i=0}^n z_i^2 = R^2 
\eeq 

\noindent embedded in a space with metric $(-, \dots ,-)$. The coordinates of 
the embedding can be re-expressed in terms of the global ones:

\bea
&AdS_n \rightarrow \left\{ \begin{array}{l} z_0 = R \cosh \rho \cos t \\ z_n 
= R \cosh \rho \sin t \\ z_i = R \sinh \rho ~\Omega_i \end{array} \dx. \\
&S_n \rightarrow \left\{ \begin{array}{l} z_0 = R \cos \theta \cos \psi \\ 
z_n = R \cos \theta \sin \psi \\ z_i = R \sin \theta ~\O_i \end{array} \dx. 
\eea

\noindent where the only restriction on the form of $\Omega_i$ is that:

\beq
\di \sum_{i=1}^{n-1} \Omega_i^2 = 1. \nn
\eeq

Notice that $0 \le \rho \le + \infty$, while $-\pi \le t, \psi, \theta \le 
\pi$; in fact we need, in order to reinterpret $t$ as the time, to unwrap it 
and use the so called global covering coordinates, where $- \infty \le t \le 
+ \infty$.



\begin{thebibliography}{99}

\bibitem{Aharony:1999ti}
O.~Aharony, S.~S.~Gubser, J.~M.~Maldacena, H.~Ooguri and Y.~Oz,
``Large N field theories, string theory and gravity'',
Phys.\ Rept.\  {\bf 323} (2000) 183
[arXiv:hep-th/9905111].

\bibitem{Blau:2001ne}
M.~Blau, J.~Figueroa-O'Farrill, C.~Hull and G.~Papadopoulos,
``A new maximally supersymmetric background of IIB superstring theory'',
JHEP {\bf 0201} (2002) 047
[arXiv:hep-th/0110242].

\bibitem{Metsaev:2001bj}
R.~R.~Metsaev,
``Type IIB Green-Schwarz superstring in plane wave Ramond-Ramond  background'',
Nucl.\ Phys.\ B {\bf 625} (2002) 70
[arXiv:hep-th/0112044].

\bibitem{Metsaev:2002re}
R.~R.~Metsaev and A.~A.~Tseytlin,
``Exactly solvable model of superstring in plane wave Ramond-Ramond  
background'',
Phys.\ Rev.\ D {\bf 65} (2002) 126004
[arXiv:hep-th/0202109].

\bibitem{Berenstein:2002jq}
D.~Berenstein, J.~M.~Maldacena and H.~Nastase,
``Strings in flat space and pp waves from N = 4 super Yang Mills'',
JHEP {\bf 0204} (2002) 013
[arXiv:hep-th/0202021].

\bibitem{Hikida:2002in}
Y.~Hikida and Y.~Sugawara,
``Superstrings on PP-wave backgrounds and symmetric orbifolds'',
JHEP {\bf 0206}, 037 (2002)
[arXiv:hep-th/0205200].

\bibitem{Gomis:2002qi}
J.~Gomis, L.~Motl and A.~Strominger,
``pp-wave / CFT(2) duality'',
JHEP {\bf 0211} (2002) 016
[arXiv:hep-th/0206166].

\bibitem{Lunin:2002fw}
O.~Lunin and S.~D.~Mathur,
``Rotating deformations of AdS(3) x S(3), the orbifold CFT and strings in  the pp-wave limit'',
Nucl.\ Phys.\ B {\bf 642} (2002) 91
[arXiv:hep-th/0206107].

\bibitem{Cho:2002zp}
J.~H.~Cho, T.~j.~Lee and S.~k.~Nam,
``Supersymmetries and Hopf duality in the Penrose limit of AdS(3) x S**3  x 
T**4'',
Nucl.\ Phys.\ B {\bf 650} (2003) 3
[arXiv:hep-th/0207252].

\bibitem{Russo:2002rq}
J.~G.~Russo and A.~A.~Tseytlin,
``On solvable models of type IIB superstring in NS-NS and R-R plane wave 
backgrounds'',
JHEP {\bf 0204} (2002) 021
[arXiv:hep-th/0202179].

\bibitem{Castellani:2000nb}
L.~Castellani and L.~Sommovigo,
``New AdS(3) x G/H compactifications of chiral IIB supergravity'',
JHEP {\bf 0007} (2000) 044
[arXiv:hep-th/0003102].

\bibitem{Cvetic:2002nh}
M.~Cvetic, H.~Lu, C.~N.~Pope and K.~S.~Stelle,
``Linearly-realised worldsheet supersymmetry in pp-wave background'',
[arXiv:hep-th/0209193].

\bibitem{Gueven:2000ru}
R.~Gueven,
``Plane wave limits and T-duality'',
Phys.\ Lett.\ B {\bf 482} (2000) 255
[arXiv:hep-th/0005061].

\bibitem{Lu:2002kw}
H.~Lu and J.~F.~Vazquez-Poritz,
``Penrose limits of non-standard brane intersections'',
Class.\ Quant.\ Grav.\  {\bf 19} (2002) 4059
[arXiv:hep-th/0204001].

\bibitem{Blau:2002mw}
M.~Blau, J.~Figueroa-O'Farrill and G.~Papadopoulos,
``Penrose limits, supergravity and brane dynamics'',
Class.\ Quant.\ Grav.\  {\bf 19} (2002) 4753
[arXiv:hep-th/0202111].

\bibitem{Das:2002ij}
S.~R.~Das and C.~Gomez,
``Realizations of conformal and Heisenberg algebras in pp-wave CFT 
correspondence'',
JHEP {\bf 0207} (2002) 016
[arXiv:hep-th/0206062].

\bibitem{Geroch:1969}
R.~Geroch,
``Limits of spacetimes'',
Commun.\ Math.\ Phys.\  {\bf 13} (1969) 180-193.

\bibitem{Figueroa-O'Farrill:1999va}
J.~M.~Figueroa-O'Farrill,
``On the supersymmetries of anti de Sitter vacua'',
Class.\ Quant.\ Grav.\  {\bf 16} (1999) 2043
[arXiv:hep-th/9902066].

\bibitem{Grisaru:1985fv}
M.~T.~Grisaru, P.~S.~Howe, L.~Mezincescu, B.~Nilsson and P.~K.~Townsend,
``N=2 Superstrings In A Supergravity Background'',
Phys.\ Lett.\ B {\bf 162} (1985) 116.

\bibitem{Casero:2002ks}
R.~Casero,
``Penrose limit of a non-supersymmetric RG fixed point'',
Nucl.\ Phys.\ B {\bf 649} (2003) 143
[arXiv:hep-th/0207221].

\bibitem{Itzhaki:2002kh}
N.~Itzhaki, I.~R.~Klebanov and S.~Mukhi,
``PP wave limit and enhanced supersymmetry in gauge theories'',
JHEP {\bf 0203} (2002) 048
[arXiv:hep-th/0202153].

\bibitem{deBoer:1999rh}
J.~de Boer, A.~Pasquinucci and K.~Skenderis,
``AdS/CFT dualities involving large 2d N = 4 superconformal symmetry'',
Adv.\ Theor.\ Math.\ Phys.\  {\bf 3} (1999) 577
[arXiv:hep-th/9904073].

\bibitem{Giveon:2003ku}
A.~Giveon and A.~Pakman,
``More on superstrings in AdS(3) x N'',
JHEP {\bf 0303} (2003) 056
[arXiv:hep-th/0302217].

\bibitem{Sevrin:1988ew}
A.~Sevrin, W.~Troost and A.~Van Proeyen,
``Superconformal Algebras In Two-Dimensions With N=4'',
Phys.\ Lett.\ B {\bf 208} (1988) 447.

\bibitem{Sevrin:1988ps}
A.~Sevrin, W.~Troost, A.~Van Proeyen and P.~Spindel,
``Extended Supersymmetric Sigma Models On Group Manifolds. 2. Current Algebras'',
Nucl.\ Phys.\ B {\bf 311} (1988) 465.

\end{thebibliography}
\end{document}